\documentclass[twocolumn,pra,showpacs,aps,superscriptaddress]{revtex4}

\usepackage{amsmath}
\usepackage{amsbsy}
\usepackage{amsthm}
\usepackage{amssymb}
\usepackage{latexsym}
\usepackage[dvips]{color,graphicx}

\begin{document}

\title{Finite-temperature properties of hard-core bosons confined on \\
one-dimensional optical lattices}

\author{Marcos Rigol}
\affiliation{Physics Department, University of California, Davis,
California 95616, USA}

\begin{abstract}
We present an exact study of the finite-temperature properties 
of hard-core bosons (HCB's) confined on one-dimensional optical 
lattices. Our solution of the HCB problem is based on the 
Jordan-Wigner transformation and properties of Slater 
determinants. We analyze the effects of the temperature on 
the behavior of the one-particle correlations, the momentum 
distribution function, and the lowest natural orbitals. 
In addition, we compare results obtained using the grand-canonical 
and canonical descriptions for systems like the ones 
recently achieved experimentally. We show that even for such 
small systems, as small as 10 HCB's in 50 lattice sites, 
there are only minor differences between the energies and 
momentum distributions obtained within both ensembles. 
\end{abstract}
\maketitle

\section{Introduction}

The study of ultracold quantum gases loaded on optical lattices
has become a very active area of experimental and theoretical 
reseach in recent years. Optical lattices enable enhancing 
interactions between atoms in weakly interacting Bose-Einstein 
condensates (BEC's) and reducing the effective dimensionality of the system.
They allow the experimental realization of strongly correlated 
bosons, well described by the Bose-Hubbard Hamiltonian 
\cite{fisher89,jaksch98}, with the consequent observation of the superfluid--Mott-insulator transition 
\cite{greiner02,wessel04}. In addition, optical 
lattices  have been used to obtain one-dimensional (1D) 
systems \cite{moritz03,tolra04} and to examine the
superfluid--Mott-insulator transition in 1D
\cite{stoferle03,batrouni02}.

Due to the strong effects of quantum fluctuations and the
possibility of obtaining exact theoretical results, 1D systems 
are a very attractive laboratory for both experiments and theory.
Theoretically, it was shown by Olshanii that in 1D 
in regimes of large scattering length, low densities, and low 
temperatures bosons behave as a gas of impenetrable particles known 
as hard-core bosons (HCB's) \cite{olshanii98}. Such a 1D gas 
(also recently called a Tonks-Girardeau gas) was  
introduced by Girardeau, who established an exact mapping between 
these strongly correlated bosons and noninteracting spinless 
fermions \cite{girardeau60}. Since then 1D HCB's have been 
extensively studied by different techniques in both homogeneous 
\cite{lieb63,lenard64,vaidya79,haldane81,korepin93,cazalilla04_1}
and harmonically trapped 
\cite{girardeau01,forrester03,gangardt04} systems.

The experimental realization of 1D HCB's followed after more 
than 40 years of the theoretical introduction of the model
\cite{paredes04,kinoshita04}, with \cite{paredes04} and 
without \cite{kinoshita04} an additional lattice along the 
1D axis. The additional 1D lattice \cite{paredes04} 
facilitates the achievement of the HCB regime with respect 
to the continuum case. It allows experimentalists to change 
the effective mass of the particles and, 
consequently, the ratio between interaction and kinetic 
energies \cite{paredes04}. Although at very 
low densities (when interparticle distances are much 
larger than the lattice spacing) HCB's on a lattice are equivalent
to HCB's in continuous space, this is not the case for arbitrary 
fillings \cite{cazalilla04}. On 1D lattices the HCB Hamiltonian 
can be mapped onto the 1D XY model of Lieb, Schulz, and Mattis 
\cite{lieb61}. For periodic systems this model has been also 
studied extensively in the literature \cite{mccoy68,vaidya78,mccoy83}. 
More recently, renewed interest has arisen on the properties 
of HCB's when additional confining potentials are introduced,
as the case relevant to experiments \cite{paredes04}. 

Remarkably, even in trapped inhomogeneous 
systems power-law behavior known from the periodic 
case is present \cite{rigol04_1}. The one-particle density
matrix exhibits a universal power-law decay with exponent 
$-1/2$ independent of the power of the 
confining potential \cite{rigol04_1}. These  
quasi-long-range one-particle correlations generate 
quasicondensates with occupations scaling proportional
to $\sqrt{N_b}$ (with $N_b$ the number of HCB's in the system)
\cite{rigol04_1}. The nonequilibrium dynamics of HCB's 
on 1D lattices has also been shown to display very 
interesting features. Quasicondensates of HCB's emerge at 
finite momentum when the system starts its free evolution 
from a pure Mott-insulating (Fock) state \cite{rigol04_2}. 
In adition, it was shown in Ref.\ \cite{rigol04_3} that in 
1D when there is no Mott insulator in the trap, the momentum 
distribution of expanding HCB's rapidly approaches that 
of noninteracting fermions \cite{rigol04_3}.

In this work we present an exact study of the finite-temperature 
properties of HCB's confined on 1D optical lattices. 
Following the spirit of Refs.\ \cite{rigol04_1,rigol04_2}, 
we develop an exact numerical approach based on the Jordan-Wigner 
transformation, which maps HCB's on a lattice onto noninteracting 
spinless fermions. We will focus on the effect of 
temperature on the off-diagonal behavior of one-particle 
correlations and related quantities like the momentum 
distribution function $n_k$ and the natural orbital 
occupations. The natural orbitals ($\phi^\eta$) are defined as the 
eigenfunctions of the one-particle density matrix ($\rho_{ij}$) 
\cite{penrose56},
\begin{equation}
\label{NatOrb}
\sum^N_{j=1} \rho_{ij}\phi^\eta_j=
\lambda_{\eta}\phi^\eta_i,
\end{equation}
and have occupations $\lambda_{\eta}$. (They resemble  
one-particle states in these strongly interacting systems.)
In dilute higher-dimensional gases, when only the lowest 
natural orbital  
(the highest occupied one) scales $\sim N_b$, it can be 
regarded as the BEC order parameter  \cite{leggett01}.
Here we will show that in 1D even at very low temperatures ($T$), 
when the energy of the system is almost identical to the one 
at $T=0$, the momentum distribution and the lowest natural orbital 
occupations can exhibit significant changes with respect to their 
values in the ground state.

The exposition is organized as follows. In Sec.\ II we describe
our exact approach to study finite-temperature systems. 
In Sec.\ III we discuss the properties of HCB's in a perfect box 
(an open system). HCB's confined in harmonic traps are analyzed in 
Sec.\ IV. Since in this work we follow a grand-canonical approach 
to study finite-temperature properties, in Sec.\ V we compare 
exact results obtained from a grand-canonical calculation with 
results obtained from a canonical one for small lattice sizes, 
like the ones recently achieved experimentally \cite{paredes04}. 
Finally, the conclusions are presented in Sec.\ VI.

\section{Exact finite-temperature approach}

In this section we detail the exact approach followed
to study the finite-temperature properties of HCB's confined 
on 1D lattices. The HCB Hamiltonian can be written as 
\begin{equation}
\label{HamHCB} H = -t \sum_{i} \left( b^\dagger_{i} b^{}_{i+1}
+ \text{H.c.} \right) + V_2 \sum_{i} x_i^2 \ n_{i },
\end{equation}
with the additional on-site constraints
\begin{equation}
\label{ConstHCB} b^{\dagger 2}_{i}= b^2_{i}=0, \quad  
\left\lbrace  b^{}_{i},b^{\dagger}_{i}\right\rbrace =1, 
\end{equation}
which avoid double or higher occupancy. The bosonic 
creation and annihilation operators at site $i$ 
are denoted by $b^{\dagger}_{i}$ and $b^{}_{i}$, respectively, 
and the local density operator by $n_i=b^{\dagger}_{i}b^{}_{i}$.
The brackets in Eq.\ (\ref{ConstHCB}) apply only to on-site 
anticommutation relations; for $i\neq j$, these operators 
commute as usual for bosons $[b^{}_{i},b^{\dagger}_{j}]=0$. 
In Eq.\ (\ref{HamHCB}), the hopping parameter is denoted by $t$
and the last term represents a harmonic trap with curvature $V_2$.

In order to exactly calculate HCB properties, we 
use the Jordan-Wigner transformation \cite{jordan28}
\begin{equation}
\label{JordanWigner} b^{\dag}_i=f^{\dag}_i
\prod^{i-1}_{\beta=1}e^{-i\pi f^{\dag}_{\beta}f^{}_{\beta}},\quad
b_i=\prod^{i-1}_{\beta=1} e^{i\pi f^{\dag}_{\beta}f^{}_{\beta}}
f_i \ ,
\end{equation}
which maps the HCB Hamiltonian onto the one of 
noninteracting spinless fermions,
\begin{eqnarray}
\label{HamFerm} H_F =-t \sum_{i} \left( f^\dagger_{i}
f^{}_{i+1} + \text{H.c.} \right)+ V_2 \sum_{i} x_i^2 \
n^f_{i },
\end{eqnarray}
where $f^\dagger_{i}$ and $f_{i}$ are the creation and
annihilation operators for spinless fermions at site 
$i$ and $n^f_{i}=f^\dagger_{i}f^{}_{i}$ is the local 
particle number operator. 

The mapping as presented above is only valid for open 
systems, as relevant for confined bosons in 
experiments \cite{paredes04,kinoshita04}. In such cases 
HCB's and fermions have exactly the same spectrum. In 
order to deal with 1D cyclic chains, with $N$ lattice sites, 
one needs to consider that
\begin{equation}
b^\dagger_{1} b^{}_{N}=-f^\dagger_{1}f^{}_{N}\ 
\exp\left( i\pi \sum^N_{\beta=1}n^f_{\beta}\right) ,
\end{equation}  
so that when the number of particles in the system 
[$\sum_{i}\langle n_i \rangle=\sum_{i}\langle n^f_i\rangle=N_b$]
is odd, the equivalent fermionic Hamiltonian satisfies periodic
boundary conditions; otherwise, if $N_b$ is even, antiperiodic 
boundary conditions are required in Eq.\ (\ref{HamFerm}).

Since for finite temperatures we will consider a 
grand-canonical ensemble---i.e., a system with fluctuating 
number of particles---in order to avoid the dependence of 
the equivalent fermionic Hamiltonian on $N_b$ we restrict 
our analysis to the open case. In this case the nontrivial 
differences between the properties of HCB's and fermions are only 
in off-diagonal correlation functions.

For finite temperatures,and within the grand-canonical formalism, 
the HCB one-particle density matrix can be written in terms 
of the equivalent fermionic system as
{\setlength\arraycolsep{0.5pt}
\begin{eqnarray}
\label{green1} \rho_{ij}&\equiv& \dfrac{1}{Z}\mathrm{Tr}\left\lbrace 
b^\dagger_{i}b^{}_{j}\exp\left[ -\left( H-\mu\sum_{l} n_l\right) /k_BT\right]\right\rbrace \nonumber \\
&=&\dfrac{1}{Z}\mathrm{Tr}\left\lbrace  f^\dagger_{i}f^{}_{j} 
\prod^{j-1}_{\beta=1} \exp (  i\pi n^f_\beta) \right. \\ 
&\times& \left. \exp\left[ -\left( H_F-\mu\sum_{l} n^f_l\right) /k_B T\right] 
\prod^{i-1}_{\gamma=1} \exp( -i\pi n^f_\gamma)  \right\rbrace ,
\nonumber
\end{eqnarray}
}where, in addition to Eqs.\ (\ref{JordanWigner}) and (\ref{HamFerm}), 
we have used the cyclic property of the trace. 
In Eq.\ (\ref{green1}), $\mu$ denotes the chemical potential, 
$k_B$ the Boltzmann constant, $T$ the temperature of the system, 
and $Z$ the partition function
\begin{equation}
\label{Z} 
Z=\mathrm{Tr} \left\lbrace 
\exp\left[-\left( H_F-\mu\sum_{l} n^f_l\right) /k_B T\right]\right\rbrace .
\end{equation}

To calculate traces over the Fock space
we will take advantage of the fact that in the equivalent 
fermionic system Fock states are Slater determinants,  
\begin{equation}
\label{wavefunct} |\Psi_{F}\rangle=\prod^{N_f}_{i=1}\ 
\sum^N_{j=1} \ P_{j i}f^{\dag}_{j}\ |0 \rangle,
\end{equation}
with $N_f$ the number of fermions and
\begin{equation}
{\bf P}=\left(
\begin{array}{c c c c c c}
P_{11}&P_{12}&\cdot&\cdot&\cdot&P_{1N_f} \\
P_{21}&P_{22}&\cdot&\cdot&\cdot&P_{2N_f} \\
\cdot&\cdot&\cdot&\cdot&\cdot&\cdot \\
\cdot&\cdot&\cdot&\cdot&\cdot&\cdot \\
\cdot&\cdot&\cdot&\cdot&\cdot&\cdot \\
P_{N1}&P_{N2}&\cdot&\cdot&\cdot&P_{NN_f} 
\end{array}
\right)
\end{equation}
the matrix of the components.

The action of exponentials bilinear on fermionic 
creation and annihilation operators, as the ones on 
Eqs.\ (\ref{green1}) and (\ref{Z}), on Slater 
determinants generates new Slater determinants
\cite{muramatsu99,assaad02}
\begin{equation}
\exp\left( \sum_{ij}f^\dagger_{i} X_{ij}f^{}_{j}\right) 
|\Psi_{F}\rangle= \prod^{N_f}_{i=1}\ \sum^N_{j=1} \ 
P'_{j i}f^{\dag}_{j}\ |0 \rangle,
\end{equation}
where
\begin{equation}
{\bf P'}=e^{\bf X}{\bf P}.
\end{equation}

Using this property one can prove the following identity 
for the trace over the fermionic Fock space \cite{muramatsu99,assaad02}
\begin{eqnarray}
&&\mathrm{Tr} \left[ \exp\left( \sum_{ij}f^\dagger_{i} X_{ij}f^{}_{j}\right) 
\exp\left(\sum_{kl}f^\dagger_{k} Y_{kl}f^{}_{l} \right) \cdots \right. \nonumber\\
&&\left. \qquad \exp\left(\sum_{mn}f^\dagger_{m} Z_{mn}f^{}_{n}\right)  \right] \nonumber \\
&&\ \ =\det\left[{\bf I}+ e^{\bf X}e^{\bf Y}\cdots e^{\bf Z}\right], 
\end{eqnarray}
which immediately allows one to calculate the partition function as
\begin{eqnarray}
\label{Z1} 
Z&=&\det\left[{\bf I}+ e^{-({\bf H}_F-\mu {\bf I})/k_B T}\right]\nonumber\\
&=&\prod_i \left[ 1+e^{-(E_{ii}-\mu)/k_B T} \right],
\end{eqnarray}
where ${\bf I}$ is the identity matrix. The last equality was obtained 
after diagonalizing Hamiltonian (\ref{HamFerm}), 
${\bf H}_F {\bf U}={\bf U E}$, ${\bf U}$ is the orthogonal matrix
of eigenvectors, and ${\bf E}$ is the diagonal matrix of eigenvalues.

The trace in Eq.\ (\ref{green1}) is calculated along the same line.
For $i\neq j$, we notice that
\begin{equation}
f^\dagger_{i}f^{}_{j}= \exp\left( \sum_{mn}f^\dagger_{m} A_{mn}f^{}_{n}\right) -1,
\end{equation}
where the only nonzero element of ${\bf A}$ is $A_{ij}=1$. Then, for 
$i\neq j$, $\rho_{ij}$ can be obtained as
{\setlength\arraycolsep{0.9pt}
\begin{eqnarray}
\rho_{ij} = \dfrac{1}{Z} &&\left\lbrace  \det \left[{\bf I}+ 
({\bf I}+{\bf A}){\bf O}_1 {\bf U} e^{-({\bf E}-\mu {\bf I})/k_B T} 
{\bf U}^\dagger {\bf O}_2\right]\right.  \nonumber \\
&&\left. -\det\left[{\bf I}+ {\bf O}_1 {\bf U} 
e^{-({\bf E}-\mu {\bf I})/k_B T}{\bf U}^\dagger {\bf O}_2\right]
\right\rbrace .
\end{eqnarray}
}${\bf O}_1$ (${\bf O}_2$) is diagonal with the first $j-1$ ($i-1$) 
elements of the diagonal equal to $-1$ and the others equal to $1$. 

The diagonal elements of the one-particle density matrix are the same of  
noninteracting fermions [see Eq.\ (\ref{green1}) for $i= j$]
and can be easily calculated as 
\cite{muramatsu99,assaad02}
\begin{eqnarray}
\rho_{ii}&=&\left[{\bf I}+ 
e^{-({\bf H}_F-\mu {\bf I})/k_B T}\right]^{-1}_{ii}  \nonumber \\
&=&\left[{\bf U} \left({\bf I}+e^{-({\bf E}-\mu {\bf I})/k_B T}\right)^{-1}  
{\bf U}^\dagger \right]_{ii}.
\end{eqnarray}
As usual, the chemical potential is fixed using the 
relation $N_b=\sum_i \rho_{ii}$ to obtain the desired
number of particles in the system.

\section{Hard-core bosons in a box}

In this section we study the finite-temperature properties of 
HCB's on a perfect box. In this case the HCB Hamiltonian can be
written as
\begin{equation}
\label{HamHCB1} H = -t \sum_{i=1}^{N-1} \left( b^\dagger_{i} b^{}_{i+1}
+ \text{H.c.} \right),
\end{equation}
with the additional on-site constraints (\ref{ConstHCB}).

The above Hamiltonian is particle-hole symmetric, like the one 
of periodic systems, under the transformation 
$h_i=b^{\dag}_i,\ h^{\dag}_i=b_i$ ($h^{\dag}_i$ and $h_i$ are 
hole creation and annihilation operators). 
The particle-hole symmetry implies that the off-diagonal elements of 
the one-particle density matrix for $N_b$ HCB's [$\rho_{ij}(N_b)$] 
and for $(N-N_b)$ HCB's 
[$\rho_{ij}(N-N_b)$] are identical. Diagonal elements satisfy 
the relation $\rho_{ii}(N_b)=1-\rho_{ii}(N-N_b)$. This leads to 
a momentum distribution function
\begin{equation}
n_k=\dfrac{1}{N}\sum_{jl} e^{-ik(x_j-x_l)}\rho_{jl},
\end{equation}
which satisfies the relation 
\begin{equation}
\label{MomHCB}
n_k(N_b)= n_{-k}(N-N_b)+\left( 1-\frac{N-N_b}{N/2}\right).
\end{equation}

\begin{figure}[h]
\includegraphics[width=0.39\textwidth]{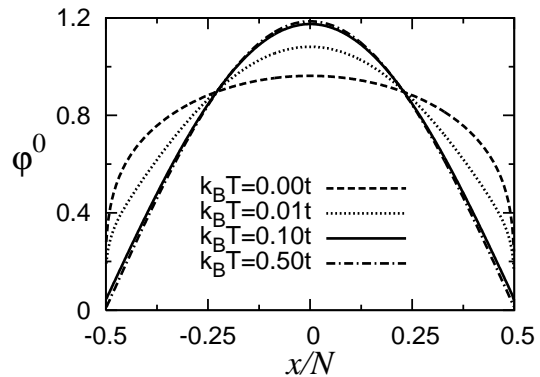}
\caption{Wave function of the lowest natural orbital at different 
temperatures in half-filled systems with 1000 lattice sites.}
\label{box_NO0}
\end{figure}

In contrast to periodic systems where the natural orbitals 
[Eq.\ (\ref{NatOrb})] are momentum states 
\cite{forrester03,rigol04_1}, this is not the case in a box. 
(The system is not translationally invariant.) In 
Fig.\ \ref{box_NO0} we show the lowest-natural-orbital wave function in a
box at different temperatures. We have normalized it as
\begin{equation}
\varphi^0=R^{1/2} \phi^0, \quad 
R=\left( N_b N\right)^{1/2}, \label{BOX_NOSGS}
\end{equation}
so that $\varphi^0$ vs $x/N$ is independent of the system size when 
the density $\rho=N_b/N$ is kept constant.

Figure \ref{box_NO0} shows that at finite temperatures the weight 
of the lowest natural orbital increases in the center of the system, 
departing from the constant value it would have in the periodic case 
($k=0$ state). Still, we find that qualitatively (and  
quantitatively) the natural orbital occupations behave very similarly
to the occupations of the momentum states so that for the box 
we will restrict our analysis to $n_k$. The natural orbitals will 
be relevant to the discussion in the harmonic trap where their
behavior can be qualitatively different to the one of $n_k$.

\begin{figure}[h]
\begin{center}
\includegraphics[width=0.49\textwidth, height=0.49\textwidth]
{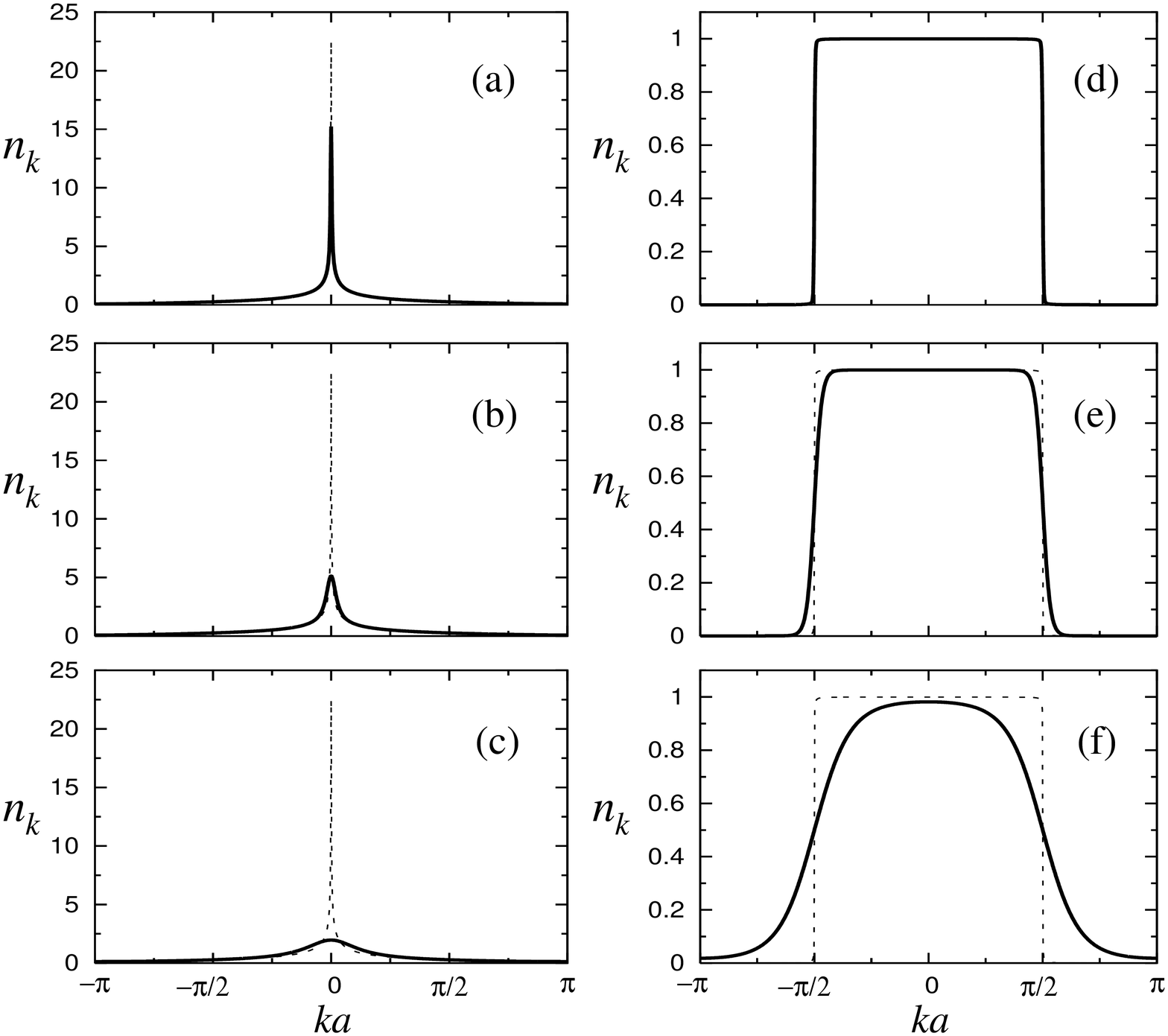}
\end{center} \vspace{-0.7cm}
\caption{Momentum distribution function of HCB's (a)--(c) and 
noninteracting fermions (d)--(f) for half-filled systems with 1000 
lattice sites at different temperatures (solid line).
The temperature [energy] of the system in each case is
$k_B T=0.01t$ [$E=-6.362\times 10^{2}t$] (a),(d), 
$0.10t$ [$-6.336\times 10^{2}t$] (b),(e),
and $0.50t$ [$-5.585\times 10^{2}t$] (c),(f). The dashed line in all 
the figures depicts the ground-state result 
($E=-6.363\times 10^{2}t$).}
\label{PerfilKBOX_Nb0500}
\end{figure}

In Figs.\ \ref{PerfilKBOX_Nb0500}(a)--\ref{PerfilKBOX_Nb0500}(c) 
we show the HCB momentum 
distribution function for half-filled systems with $N=1000$ 
and different temperatures. We have plotted as dashed lines 
the ground-state results for comparison. The effects of small 
but finite temperatures are dramatic. This can be better seen 
in Figs.\ \ref{PerfilKBOX_Nb0500}(a) and \ref{PerfilKBOX_Nb0500}(b) 
where the energies of the 
finite-temperature systems are almost identical to the ones of 
the ground state. For $k_B T=0.01t$, the relative energy 
difference $(\delta E= [E(T)-E(0)]/|E(0)|)$ between the 
finite-temperature system [$E(T)$] and the ground state [$E(0)$] is 
$\delta E\sim 0.01\%$. In Fig.\ \ref{PerfilKBOX_Nb0500}(a) one can 
see that the $k=0$ momentum peak is already around 2/3 of the one 
at zero temperature. For the case in Fig.\ \ref{PerfilKBOX_Nb0500}(b), 
$\delta E\sim 0.4\%$ and the peak at $n_{k=0}$ has already reduced 
almost 5 times. At $k_B T=0.5t$ in Fig.(a), the zero momentum peak 
has practically disappeared.

As opposed to the HCB momentum distribution function, we have
plotted in Figs.\ \ref{PerfilKBOX_Nb0500}(d)--\ref{PerfilKBOX_Nb0500}(f) 
the momentum 
distribution function of the equivalent noninteracting fermions.
These figures not only show the differences between the shape 
of the momentum distributions in both cases, but also the fact 
that they are affected very differently by the temperature.
In the fermionic case it is well known that the changes on 
$n_k$ occur only around the Fermi surface and are of order 
$k_B T$, so that in Fig.\ \ref{PerfilKBOX_Nb0500}(d) one cannot 
notice the differences between the finite- and zero-temperature 
cases. In Fig.\ \ref{PerfilKBOX_Nb0500}(e) they are very small, 
and only when $k_B T$ becomes of the order of $t$ 
[Fig.\ \ref{PerfilKBOX_Nb0500}(f)] can one see a large deviation 
of the finite-temperature $n_k$ with respect to the one in the 
ground state.

The zero-temperature peaks in the HCB $n_k$ 
[Figs.\ \ref{PerfilKBOX_Nb0500}(a)--Figs.\ \ref{PerfilKBOX_Nb0500}(c)] 
reflect the presence of quasi-long-range one-particle correlations 
\cite{mccoy68,vaidya78,mccoy83,rigol04_1}; i.e., there is a 
power-law decay $\rho_{ij}\sim |x_i -x_j|^{-1/2}$. In these 
1D systems any finite temperature generates an exponential
decay of $\rho_{ij}$, which destroys the quasi-long-range 
correlations present in the ground state. 
This exponential decay is the one producing dramatic effects 
in $n_k$. 

\begin{figure}[h]
\includegraphics[width=0.39\textwidth]{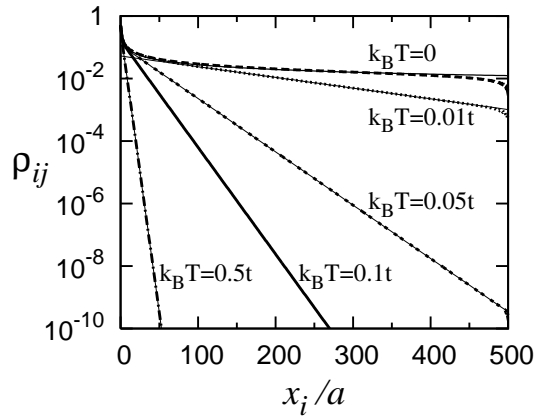}
\caption{Decay of the one-particle density matrix at different 
temperatures in half-filled 
systems with 1000 lattice sites. Thin solid lines following 
finite-temperature results exhibit exponential decays, while the one 
following the $T=0$ result exhibits a power-law decay $\sim x_i^{-1/2}$.
In all cases we measured $\rho_{ij}$ fixing $j$ in the middle of the 
box---i.e., $x_j=0$}
\label{box_OPDM}
\end{figure}

In Fig.\ \ref{box_OPDM} we show the decay of one-particle correlations
for the same systems of Fig.\ \ref{PerfilKBOX_Nb0500}. At very low 
temperatures ($k_B T=0.01t$) the one-particle density matrix follows the ground-state result
over a certain distance, which reduces with increasing the temperature, 
approximately up to the point where the exponential decay sets in. Our results 
in Fig.\ \ref{box_OPDM} can be compared with the ones obtained by other 
means for $\left\langle S^x_i S^x_j\right\rangle$ in the 
1D spin-1/2 isotropic XY model \cite{tonegawa81}, 
to which HCB's can be mapped. Apart from a (1/2) normalization factor 
the results agree.

\begin{figure}[h]
\includegraphics[width=0.415\textwidth]{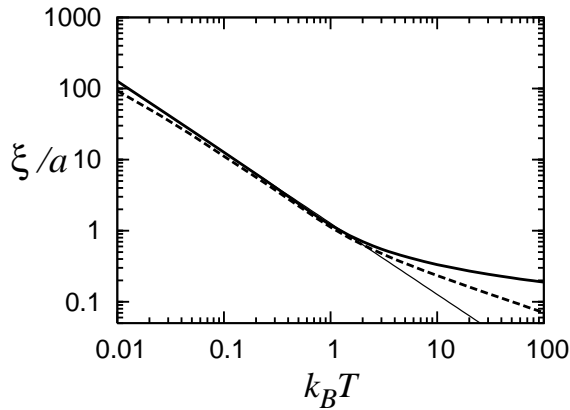}
\caption{Correlation length vs temperature (in units of $t$) in half-filled 
systems with 1000 lattice sites. We have plotted $\xi$ as a thick 
solid line and the second moment of the one-particle density matrix $\tilde{\xi}$ 
as a dashed line (see text). The thin solid line is the result 
of a fit $\xi/a=1.27t/k_B T$.}
\label{box_CorrLvsT}
\end{figure}

The quantity of relevance to characterize the finite-temperature 
exponential decay of the one-particle density matrix 
$\rho_{ij}\sim e^{-|x_i-x_j|/\xi}$ 
(Fig.\ \ref{box_OPDM}) is the correlation length $\xi$. This 
quantity is of experimental importance since for $\xi\gtrsim N$ the HCB 
gas (essentially) exhibits at finite temperatures properties 
of the ground state. At low temperatures, $k_B T<t$, the correlation 
length decreases as $\xi \sim 1/T$ with increasing 
temperature. This is shown in Fig.\ \ref{box_CorrLvsT}.

\begin{figure}[b]
\begin{center}
\includegraphics[width=0.395\textwidth]
{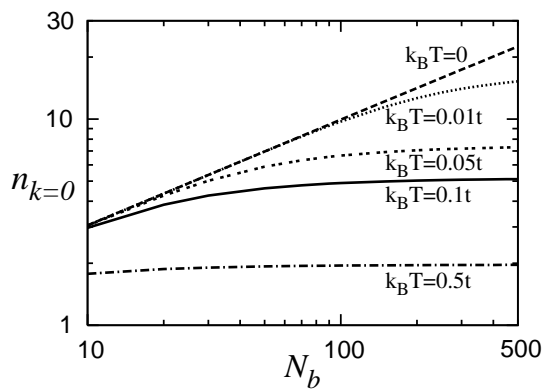}
\end{center} \vspace{-0.6cm}
\caption{Scaling of $n_{k=0}$ vs $N_b$ for a constant density 
$\rho=0.5$ and the same values of the temperature chosen in 
Fig.\ \ref{box_OPDM}.}
\label{BOX_Scaling}
\end{figure}

A way of seeing the effects that a finite-temperature 
correlation length produces in these bosonic systems is
to study how the occupation of the zero-momentum state scales 
with the number of particles (or the system size) when the density 
is kept constant. Results for $n_{k=0}$ vs $N_b$ are 
presented in Fig.\ \ref{BOX_Scaling}. There we have plotted results 
for as many temperatures as in Fig.\ \ref{box_OPDM} so that 
one can see at what system size the finite-temperature results 
depart from the ones of the ground state. Since the system size 
is twice the number of particles (they are at half filling), 
one can then notice, with the help of Fig.\ \ref{box_CorrLvsT}, 
that the mentioned departure indeed occurs for system sizes larger 
than the correlation length. For example, a half-filled box 
with 20 HCB's (a filling similar to the one achieved experimentally 
in Ref.\ \cite{paredes04}) would have a momentum distribution 
function very similar to the one in the ground state up to a 
temperature $k_B T=0.05t$. At zero temperatures $n_{k=0}$ scales
proportionally to $\sqrt{N_b}$; i.e, it diverges when 
$N_b \rightarrow \infty$, reflecting the power-law decay
of one-particle correlations shown in Fig.\ \ref{box_OPDM}
\cite{rigol04_1}.

\begin{figure}[h]
\begin{center}
\includegraphics[width=0.39\textwidth]
{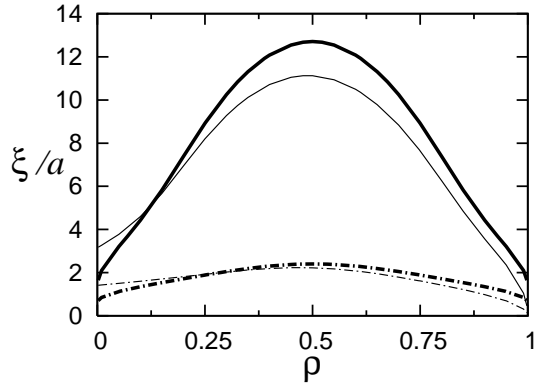}
\end{center} \vspace{-0.6cm}
\caption{Correlation length vs density for systems with $N=500$, 
$k_B T=0.1t$ (solid line) and $k_B T=0.5t$ 
(dash-dotted line). Thick lines depict $\xi$ and
thin lines the second moment of the one-particle density 
matrix $\tilde{\xi}$.}
\label{BOX_CorrLvsRho}
\end{figure}

The one-particle correlation length not only depends strongly 
on the temperature, but also on the density in the system. 
(In Fig.\ \ref{box_CorrLvsT} we have only shown results 
for the half-filled case.) The dependence of the correlation 
length on the density, for two values of the temperature, 
is depicted in Fig.\ \ref{BOX_CorrLvsRho}. Notice that both 
curves are symmetric with respect to $\rho=0.5$ due to  
particle-hole symmetry.

The strong dependence of the correlation length on the density
represents a difficulty for defining $\xi$ in inhomogeneous 
systems, like the ones achieved experimentally where HCB's are 
trapped in harmonic confining potentials. 
(In a box the density is not exactly constant, away from 
half and integer fillings, due to Friedel oscillations, 
but they reduce with increasing system size.) An alternative 
definition to the correlation length $\xi$ may be given as the 
second moment of the one-particle density matrix
\begin{equation}
\label{secmom}
\tilde{\xi}=\sqrt{\dfrac{1}{2}
\dfrac{\sum_{ij}\left( x_i-x_j \right)^2 \rho_{ij}}{\sum_{ij} \rho_{ij}}}.
\end{equation}

In Fig.\ \ref{box_CorrLvsT} we have plotted $\tilde{\xi}$ along with
$\xi$. When $1<\xi\ll N$ both $\tilde{\xi}$ and
$\xi$ are very similar. For the lowest temperatures, in 
Fig.\ \ref{box_CorrLvsT}, we considered systems with 1000 lattice 
sites, which are not much larger than the correlation length. This 
is the origin of the differences between $\tilde{\xi}$ and $\xi$
observed for large values of $\xi$. At high temperatures ($k_B T>t$) 
the value of $\tilde{\xi}$ is completely dominated by the 
very-short-distance sector of the one-particle density matrix, 
so that $\tilde{\xi}$ and $\xi$ are 
expected to be very different. At intermediate temperatures one 
can use $\tilde{\xi}$ as a good estimate of $\xi$. 

In Fig.\ \ref{BOX_CorrLvsRho} we have also plotted $\tilde{\xi}$ 
along with $\xi$ so that one can realize how the inclusion of 
the short-range part of the one-particle density matrix 
in $\tilde{\xi}$ produces different 
effects for low densities, where $\tilde{\xi}>\xi$, and high densities,
where $\tilde{\xi}<\xi$. Still the overall behavior of $\tilde{\xi}$ 
is similar to the one of $\xi$. In the next section we will rely on 
$\tilde{\xi}$ for estimating the correlation length in harmonic traps 
and also for comparing it to the one in the box.

\section{Hard-core bosons in harmonic traps}

We study in this section HCB's trapped in harmonic potentials. 
The addition of a confining potential generates a 
position-dependent density profile where, at zero temperature, 
superfluid and Mott-insulating regions can coexist. 
In the next two subsections we analyze the effects 
of the temperature on density and 
momentum profiles of systems in which the ground state is
(i) superfluid (Sec.\ \ref{superf}) and (ii) a coexistence 
of superfluid and Mott-insulating phases (Sec.\ \ref{Mott}). 
In Sec.\ \ref{TRAP_Corr} we address more general questions 
like the behavior of one-particle correlations 
and scaling properties at finite temperatures.

In harmonic traps we normalize $n_k$ 
using a length scale set by the combination lattice-confining 
potential, 
\begin{equation}
\zeta=\left(V_{2}/t \right)^{-1/2},
\end{equation}
so that 
\begin{equation}
n_k=\dfrac{a}{\zeta}\sum^N_{jl=1} e^{-ik(j-l)} \rho_{jl}.
\end{equation}
In addition, instead of the density $\rho=N_b/N$, 
relevant to the periodic or open case, we consider the 
characteristic density \cite{rigol04_1,rigol03_3}
\begin{equation}
\tilde{\rho}=N_b a/\zeta.
\end{equation}
As shown in Ref.\ \cite{rigol03_3} up to 
$\tilde{\rho}\sim 2.6$--2.7 there is no Mott insulator in the
trap. For larger values of $\tilde{\rho}$ a Mott-insulating
phase appears in the middle of the system.

\subsection{Superfluid case at $T=0$ \label{superf}}

\begin{figure}[h]
\begin{center}
\includegraphics[width=0.49\textwidth,height=0.52\textwidth]
{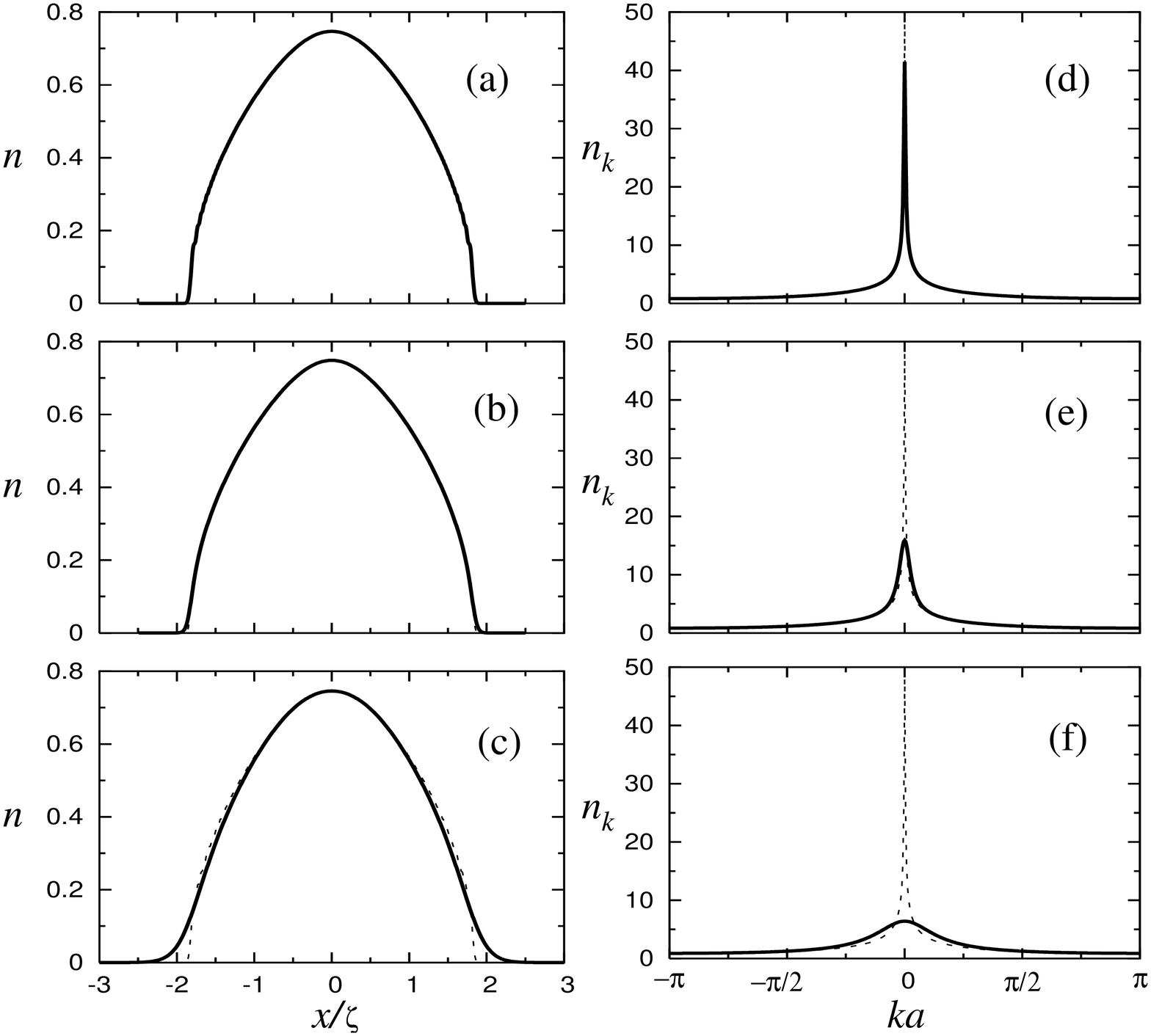}
\end{center} \vspace{-0.5cm}
\caption{Density (a)--(c) and normalized momentum distribution function 
(d)--(f) of 200 HCB's in a trap with $V_2 a^2= 10^{-4}t$ 
($\tilde{\rho}=2$) at different temperatures (solid line). 
The temperature [energy] of the system in each case is  
$k_B T=0.01t$ [$E=-37.71t$] (a),(d), $0.10t$ [$-36.44t$] (b),(e), 
and $0.50t$ [$-8.48t$] (c),(f). The dashed line in all the figures
depicts the ground-state result ($E=-37.72t$).}
\label{Perfiles_Nb0200}
\end{figure}

In Fig.\ \ref{Perfiles_Nb0200} we show density and momentum 
profiles in a trap with 200 HCB's ($\tilde{\rho}=2$) for 
different temperatures (solid line) and compared to the 
ground-state case (dashed line). As for the fermionic $n_k$
in Figs.\ \ref{PerfilKBOX_Nb0500}(d)--\ref{PerfilKBOX_Nb0500}(f), 
the changes of the density profiles with the temperature in 
Figs.\ \ref{Perfiles_Nb0200}(a)--\ref{Perfiles_Nb0200}(c) 
are the ones expected for fermions. 
[HCB's and fermions exhibit identical density profiles, Eq.\ (\ref{green1}).] 
For temperatures much smaller than the Fermi energy, which is of the 
order of $t$ for these systems, the density profiles almost do
not change. The same occurs with the total energy $E$ of the 
trapped cloud, as seen from their values reported in the caption
of Fig.\ \ref{Perfiles_Nb0200}. On the other hand, the behavior
of $n_k$, related to off-diagonal one-particle correlations, is
very different to the one of the density. At $k_B T=0.01t$ 
[Figs.\ \ref{Perfiles_Nb0200}(a) and \ref{Perfiles_Nb0200}(d)], 
when the energy of the 
system has changed by $0.03\%$ with respect to the ground-state 
energy, changes can be already noticed in $n_k$ around $k=0$. 
For $k_B T=0.1t$ [Figs.\ \ref{Perfiles_Nb0200}(b) and 
\ref{Perfiles_Nb0200}(e)], 
the energy is $3\%$ larger than in the ground state 
and the peak in $n_{k=0}$ is less than one-third of its value at $T=0$. 
For larger temperatures, like $k_B T=0.5t$ in 
Figs.\ \ref{Perfiles_Nb0200}(c) and \ref{Perfiles_Nb0200}(f),
almost no peak can be seen 
in $n_{k=0}$ as compared with the one in the ground state.
This is similar to the results obtained for the box in the 
previous section, with a difference being that in the box the 
density distribution is not affected by the temperature.

\begin{figure}[h]
\begin{center}
\includegraphics[width=0.49\textwidth,height=0.52\textwidth]
{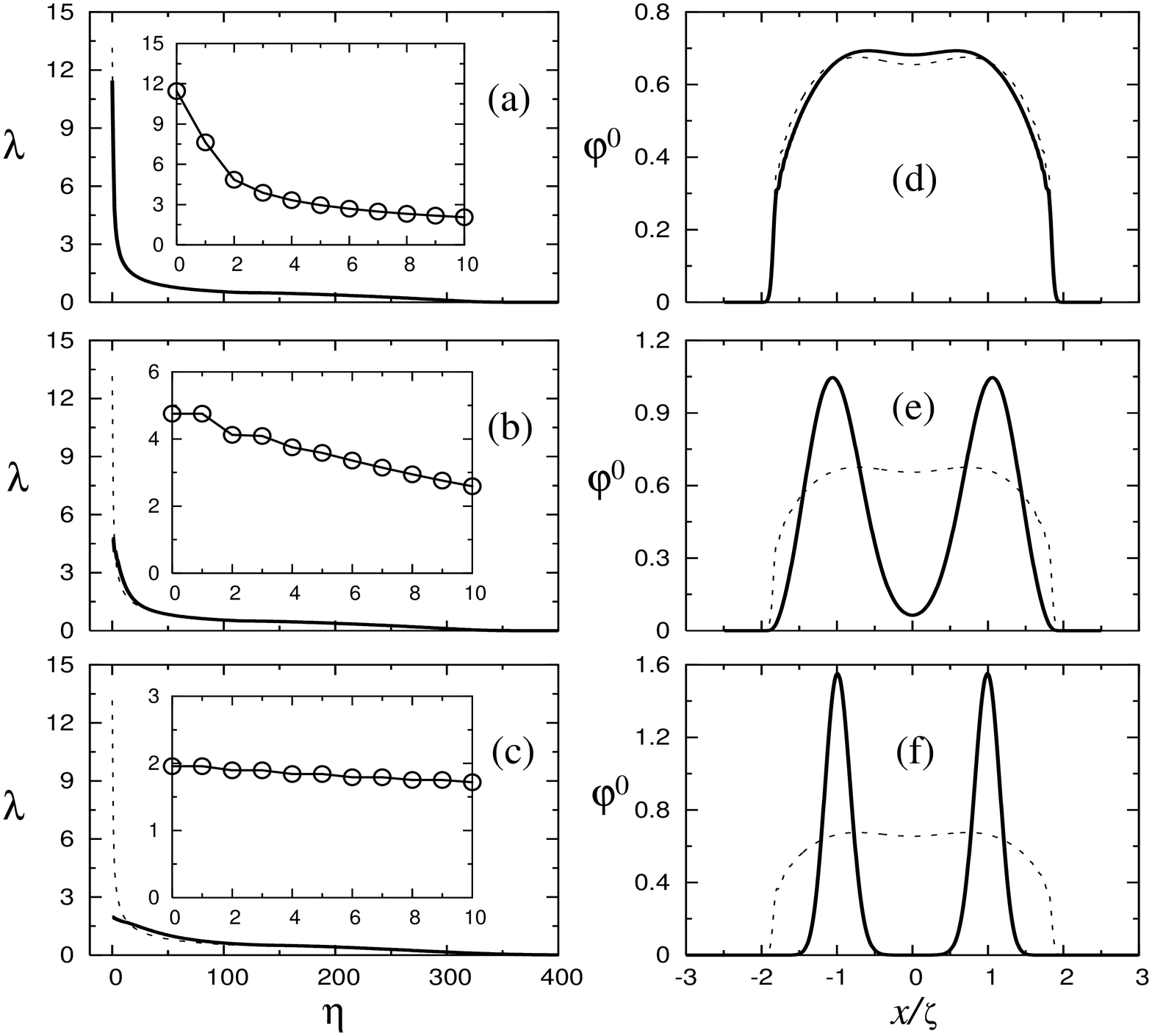}
\end{center} \vspace{-0.5cm}
\caption{Natural orbital occupations (a)--(c) and normalized 
wave function of the lowest natural orbital (d)--(f) for the 
same systems shown in 
Fig.\ \ref{Perfiles_Nb0200}, $N_b=200$ and 
$V_2 a^2=10^{-4}t$ ($\tilde{\rho}=2$). 
The temperature [energy] of the system in each case
(solid line) is  $k_B T=0.01t$ [$E=-37.71t$] (a),(d), 
$0.10t$ [$-36.44t$] (b),(e), and $0.50t$ [$-8.48t$] (c),(f). 
The dashed line in all the figures depicts the ground-state 
result ($E=-37.72t$). The insets in (a)--(c) display 
in more detail the lowest 11 natural orbital occupations.}
\label{NatOrb_Nb0200}
\end{figure}

Other quantities of relevance to the harmonically trapped 
case are the natural orbital occupations 
[Figs.\ \ref{NatOrb_Nb0200}(a)--\ref{NatOrb_Nb0200}(c)] 
and the wave function of the lowest natural orbital 
[Figs.\ \ref{NatOrb_Nb0200}(d)--\ref{NatOrb_Nb0200}(f)]. 
In Figs.\ \ref{NatOrb_Nb0200}(d)--\ref{NatOrb_Nb0200}(f) 
we normalize the natural orbital wave function following 
Ref.\ \cite{rigol04_3}
\begin{equation}
\varphi^0=R^{1/2} \phi^0, \quad 
R=\left( N_b \zeta/ a\right)^{1/2}. \label{NOSGS}
\end{equation}
Like $n_k$, the natural orbital occupations exhibit a very strong 
dependence on the temperature, which can be understood since they are
also related to the off-diagonal one-particle correlations.
More interesting, and qualitatively different to the case in the 
box, is the behavior displayed by the lowest-natural-orbital wave 
function in Fig.\ \ref{NatOrb_Nb0200}. With increasing temperature, 
for large fillings, the weight of the lowest 
natural orbital in the middle of the trap decreases, and for 
$k_B T=0.5t$ [Fig.\ \ref{NatOrb_Nb0200}(f)] it is exactly zero. 
This behavior of the wave function is accompanied by the 
appearance of a degeneracy in the occupation of the lowest 
natural orbitals. These two effects are very similar to the 
ones generated by the increase of the filling in the ground 
state of the system and the formation of a Mott 
insulator in the middle of the trap \cite{rigol04_1}. 
However, as seen in 
Figs.\ \ref{Perfiles_Nb0200}(a)--\ref{Perfiles_Nb0200}(c) 
no Mott insulator is created by an increase of the 
temperature.

In order to understand the above effect it is important to realize
that the spectrum of noninteracting particles in a combination
lattice-harmonic potential \cite{rigol03_3,hooley04,ruuska04,rey05} 
is very different to the one of the harmonic oscillator in the 
continuum. (In the latter case one can intuitively realize that 
the maximum weight of a condensate, or of the largest eigenvalue 
of the one-particle density matrix, occurs in the middle of the trap.) 
In a lattice with a superposed 
harmonic oscillator, the eigenvalues of the noninteracting Hamiltonian 
reduce their weight in the middle of the system when their energy 
increases. For energies larger of $2t$ 
[for the Hamiltonian in Eq.\ (\ref{HamFerm})] \cite{BW}, 
the eigenstates of the Hamiltonian start to be localized at the sides
of the trap; i.e., they have zero weight in the center of the system
\cite{rigol03_3,hooley04,ruuska04,rey05}.
At zero temperatures a Mott-insulating domain in the center 
of the trap signals that these states are populated \cite{rigol03_3}.
At finite temperatures the occupation of localized states occurs
even when there is no Mott insulator in the system, 
which explains why the lowest natural orbital can exhibit a behavior 
like the one seen in Figs.\ \ref{NatOrb_Nb0200}(e) 
and \ref{NatOrb_Nb0200}(f) in the absence of the insulating core.

Before analyzing the finite-temperature one-particle correlations 
in the confined system, which explain the previously observed 
effects in $n_k$ and the natural orbital occupations, we present in 
what follows an example of the consequences of the temperature in a 
system that in its ground state exhibits a coexistence of superfluid 
and Mott-insulating phases.

\subsection{Mott insulator is present at $T=0$ \label{Mott}}

\begin{figure}[h]
\begin{center}
\includegraphics[width=0.49\textwidth,height=0.34\textwidth]
{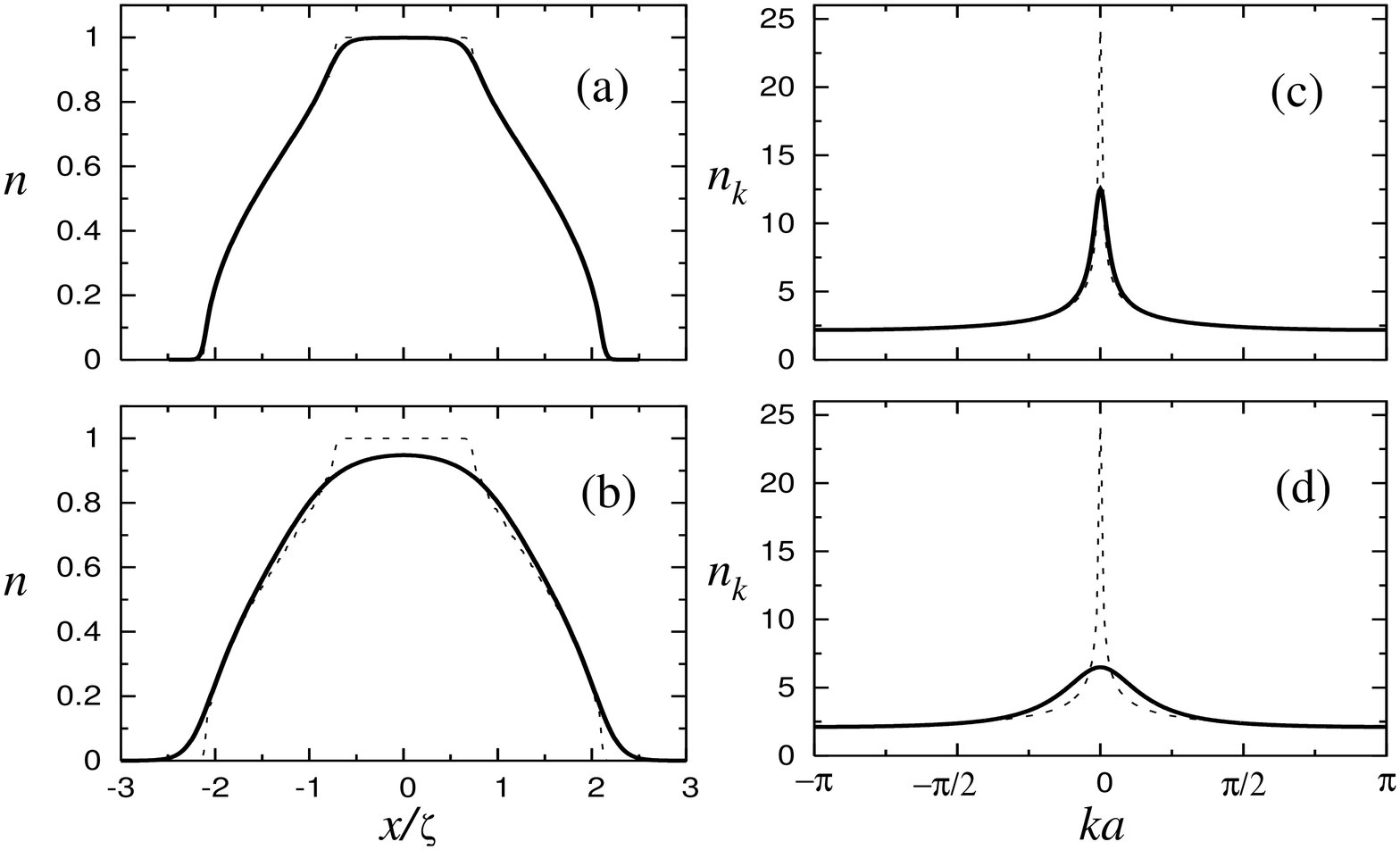}
\end{center} \vspace{-0.5cm}
\caption{Density (a),(b) and normalized momentum distribution function 
(d),(d) of 300 HCB's in a trap with $V_2 a^2= 10^{-4}t$ 
($\tilde{\rho}=3$) at different temperatures (solid line). 
The temperature [energy] of the system in each case is  
$k_B T=0.1t$ [$E=158.7t$] (a),(c) and $0.5t$ [$185.9t$] (b),(d). 
The dashed line in all the figures depicts the ground-state result 
($E=157.4t$).}
\label{Perfiles_Nb0300}
\end{figure}

In Fig.\ \ref{Perfiles_Nb0300} we show density and momentum 
profiles of a system with 300 HCB's ($\tilde{\rho}=3$) for 
different temperatures (solid line) and compared to the 
ground-state case (dashed line). As for the superfluid case
discussed in the previous subsection, density profiles are
almost not modified for $k_B T\ll t$. Increasing the 
temperature one can see in Fig.\ \ref{Perfiles_Nb0300}(b)
that as $k_B T$ approaches $t$ the Mott insulating 
($n=1$) plateau in the middle of the trap disappears. 
The effects of the temperature in $n_k$ are also similar 
to the ones in the case with no Mott insulator. $n_k$ 
strongly depends on the temperature in the system.

It is worth noticing in Figs.\ \ref{Perfiles_Nb0300}(c) and 
\ref{Perfiles_Nb0300}(d)
that even in the presence of a Mott-insulating phase, at zero 
temperature, $n_k$ exhibits a sharp peak at $k=0$ due to 
the superfluid phases at the sides 
\cite{wessel04,rigol04_1}. The effects 
of the Mott insulator in $n_k$ are reflected by a large population 
of $k$ states around $ka=\pm \pi$ and an increase of the 
full width at half maximum of the $k=0$ peak. These are characteristics 
of the system that remain at finite but very low temperatures. Increasing
the temperature [Fig.\ \ref{Perfiles_Nb0300}(d)] the peak at $k=0$ 
disappears. On the other hand, the high-momentum tails remain 
almost unmodified with respect to the ground-state case as they 
reflect the properties of short-distance correlations, related to the 
density profiles, which are much less sensitive to temperature effects.

\begin{figure}[h]
\begin{center}
\includegraphics[width=0.49\textwidth,height=0.34\textwidth]
{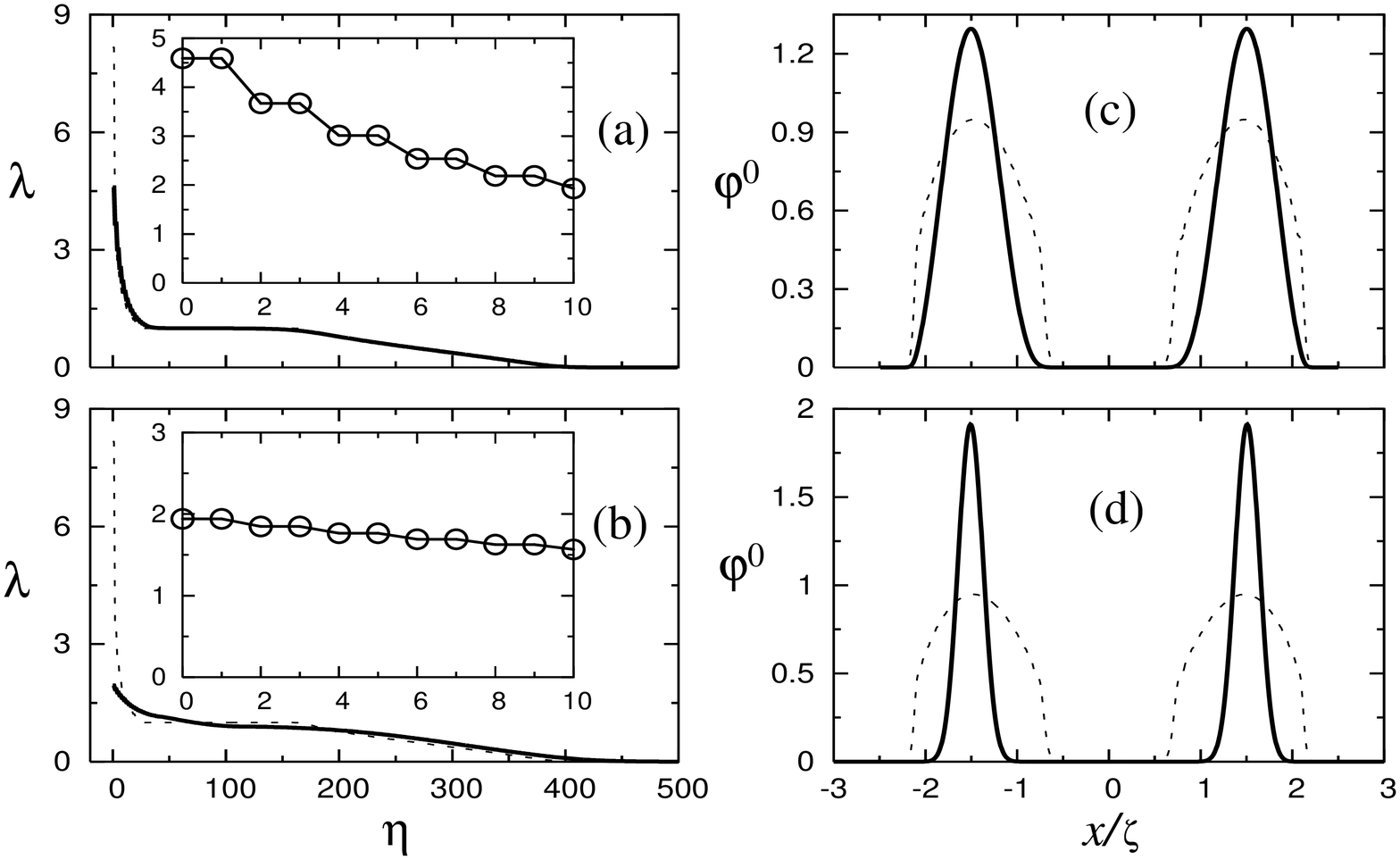}
\end{center} \vspace{-0.7cm}
\caption{Natural orbital occupations (a),(b) and normalized wave 
function of the lowest natural orbital (c),(d) for the same systems 
shown in Fig.\ \ref{Perfiles_Nb0300}, $N_b=300$ and 
$V_2 a^2= 10^{-4}t$ ($\tilde{\rho}=3$). 
The temperature [energy] of the system in each case
is  $k_B T=0.1t$ [$E=158.7t$] (a),(c) 
and $0.5t$ [$185.9t$] (b),(d). The dashed line in all the 
figures depicts the ground-state result ($T=0$ and $E=157.4$). 
The insets in (a),(b) display in more detail the lowest 
11 natural orbital occupations.}
\label{NatOrb_Nb0300}
\end{figure}

At zero temperature, the natural orbital occupations exhibit 
a clear signature of the presence of the Mott-insulating 
core in the trap. A plateau with $\lambda_\eta=1$ is present, reflecting 
the existence of single occupied states. In this case the lowest natural 
orbital is degenerate due to the splitting of the system by the 
Mott-insulating core. Two 
identical quasicondensates can be observed at the sides of the Mott 
core [Figs.\ \ref{NatOrb_Nb0300}(c) and \ref{NatOrb_Nb0300}(d)]. 
The increase of temperature reduces the occupation of the lowest 
natural orbital [they become more localized, 
Figs.\ \ref{NatOrb_Nb0300}(c) and \ref{NatOrb_Nb0300}(d)], 
but does not destroy their degeneracy. This degeneracy in absence 
of a Mott-insulating state is, as explained in the previous 
subsection, an effect that only appears at finite temperatures 
due to the population of localized states at the sides of the trap.
Finally, one should notice that the plateau with $\lambda_\eta=1$
disappears in Fig.\ \ref{NatOrb_Nb0300}(b) along with the 
disappearance of the Mott plateau in Fig.\ \ref{Perfiles_Nb0300}(b).

\subsection{Correlation functions and scalings \label{TRAP_Corr}}

\begin{figure}[h]
\begin{center}
\includegraphics[width=0.48\textwidth,height=0.19\textwidth]
{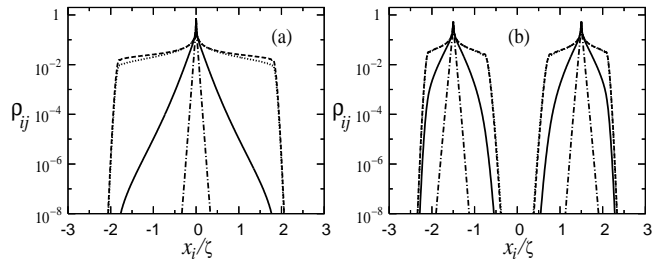}
\end{center} \vspace{-0.6cm}
\caption{One-particle correlations corresponding to the systems 
analyzed in Secs.\ \ref{superf} and \ref{Mott}. (a) $N_b=200$,  
$V_2 a^2= 10^{-4}t$ ($\tilde{\rho}=2$), and
$x_j/\zeta=0$. (b) $N_b=300$, $V_2 a^2= 10^{-4}t$ 
($\tilde{\rho}=3$), and $x_j/\zeta=\pm 1.5$. Different 
temperatures are denoted by $k_B T=0$ (dashed line),
$0.01t$ (dotted line), $0.10t$ (solid line), and
$0.50t$ (dash-dotted line).}
\label{TRAP_OPDM}
\end{figure}

In Fig.\ \ref{TRAP_OPDM} we show the behavior of one-particle
correlations with increasing temperature for the two cases 
analyzed in the previous subsections. Correlations ($\rho_{ij}$) 
are measured with respect to a fixed point $x_j$, while
$x_i$ is changed all over the system. In Fig.\ \ref{TRAP_OPDM}(a)
the correlations are measured with respect to the middle of the 
trap and in Fig.\ \ref{TRAP_OPDM}(b) with respect to two points
at the sides of the Mott-insulating core present at $T=0$.

In the ground state, the one-particle density matrix decays as a power law 
$\rho_{ij}\sim|x_i-x_j|^{-1/2}$ for $0< n_i,n_j < 1$ \cite{rigol04_1}. 
The introduction of a small temperature $k_B T=0.01t$ can be already 
noticed in Fig.\ \ref{TRAP_OPDM}(a) as a faster decay of correlations 
at long distances. At temperatures larger than $k_B T=0.1t$, 
for the system sizes of the figure, the one-particle density matrix decays exponentially 
to $10^{-8}$ before reaching the borders of the trap. Due to the space 
varying density one can notice that in contrast to the box, 
in a harmonic trap one cannot see a single correlation length, 
which would mean a straight line in all the semilogarithmic plots of 
the figure. Still one can calculate the second moment of the 
one-particle density matrix $\tilde{\xi}$ [Eq.\ (\ref{secmom})]
as a sort of an averaged correlation length.

\begin{table}[h]
\caption{Second moment of the one-particle density matrix 
($\tilde{\xi}$), in harmonic traps ($V_2 a^2=10^{-4} t$), 
for different temperatures, and characteristic 
densities. We also present values of $\tilde{\xi}$ 
calculated in boxes ($N=500$) with the same densities than 
in the center of the trapped case. Notice that in contrast 
to the characteristic density, the density in the center 
of the trap changes with increasing temperature. 
The two values of $\rho$ correspond to 
$k_B T=0.1t$ and $0.5t$, respectively.}
\begin{tabular}{cccc}
\hline\hline
$\tilde{\xi}/a$ (trap)&$k_B T=0.1t$&$k_B T=0.5t$&$\tilde{\xi}/a$ {\it (box)} \\
\hline 
$\tilde{\rho}=0.5$ &8.7 /{\it 9.7}&1.9 /{\it 2.0}&$\rho=${\it 0.33,0.29}   \\
$\tilde{\rho}=1.0$ &10.4 /{\it 11.1}&2.1 /{\it 2.2}&$\rho=${\it 0.48,0.47} \\
$\tilde{\rho}=1.5$ &10.5 /{\it 10.3}&2.1 /{\it 2.1}&$\rho=${\it 0.61,0.61} \\
$\tilde{\rho}=2.0$ &9.7 /{\it 7.7}&2.0 /{\it 1.6}&$\rho=${\it 0.75,0.75} \\
$\tilde{\rho}=2.5$ &8.8 /{\it 2.8}&1.8 /{\it 1.1}&$\rho=${\it 0.93,0.87} \\
$\tilde{\rho}=3.0$ &8.5 /{\it 0.0}&1.7 /{\it 0.7}&$\rho=${\it 1.00,0.95} \\
\hline\hline
\end{tabular}
\label{TRAPvsBOXM}
\end{table}

We present in Table \ref{TRAPvsBOXM} results for $\tilde{\xi}$ 
in harmonic traps for two temperatures and six values of 
$\tilde{\rho}$. To the right we show results obtained in boxes 
with densities chosen to be identical to the ones at 
the center of the harmonically trapped cloud. One can see 
that the results in both cases are similar far from the region 
where the Mott insulator sets in the middle of the trap 
($\tilde{\rho}=$0.5--2.0 in Table \ref{TRAPvsBOXM}), 
so that one can estimate $\tilde{\xi}$ in harmonic traps 
using results from a box. This is in agreement with recent 
results reported for other finite-temperature correlation 
lengths in trapped bosonic systems with no lattice 
\cite{kheruntsyan05}. The reason for the agreement between 
$\tilde{\xi}$ in the trap and in the box is that in the
first case $\tilde{\xi}$ is dominated by the contributions 
of the middle of the system, where the density has its maximum 
value and it is ``relatively uniform.'' As one approaches 
$n=1$ in the middle of the trap, or $\tilde{\rho}=2.6$, 
the argument above fails ($\tilde{\rho}=2.5,3.0$ in 
Table \ref{TRAPvsBOXM}) since the correlation 
length in the center of the cloud approaches zero 
(see $\tilde{\xi}$ vs $\rho$ in Fig.\ \ref{BOX_CorrLvsRho} 
when $\rho\rightarrow 1$) and regions with smaller densities 
start to dominate the value of $\tilde{\xi}$. 
For those cases an exact calculation of $\tilde{\xi}$, 
given the density profile, is required.

\begin{figure}[h]
\begin{center}
\includegraphics[width=0.38\textwidth]
{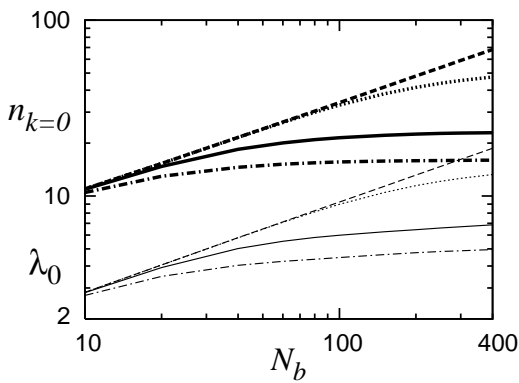}
\end{center} \vspace{-0.7cm}
\caption{Scaling of $n_{k=0}$ (thick lines) and the lowest 
natural orbital occupation (thin lines) vs $N_b$ for a 
constant characteristic density $\tilde{\rho}=2$. Different 
temperatures are plotted with the same convention of 
Fig.\ \ref{TRAP_OPDM}, $k_B T=0$ (dashed line),
$0.01t$ (dotted line), $0.05t$ (solid line), and
$0.10t$ (dash-dotted line).}
\label{TRAP_Scaling}
\end{figure}

The exponential decay of the one-particle density matrix 
implies that when the size of the system is larger than 
the averaged correlation length, the momentum distribution 
function and the occupation of the lowest
natural orbital stop changing with increasing system size. 
The size at which this
occurs depends on the temperature, as the averaged correlation 
length decreases with increasing the temperature, and also 
depends on the 
characteristic density. In Fig.\ \ref{TRAP_Scaling}
we show how $n_{k=0}$ and $\lambda_0$ scale at four different 
temperatures and starting from small system sizes (close to the 
ones achieved experimentally \cite{paredes04}). At $T=0$ the 
increase of both quantities is $\sim \sqrt{N_b}$, reflecting 
quasi-long-range correlations present in $\rho_{ij}$ \cite{rigol04_1}.
For $k_B T=0.01t$, the departure from the zero-temperature
values occurs when the trap has $\sim 100$ HCB's. For $k_B T=0.05t$ it 
occurs around $N_b=20$, and for $k_B T=0.1t$ even the smallest 
system with 10 HCB's is different to the ground state.

\begin{figure}[h]
\begin{center}
\includegraphics[width=0.48\textwidth]
{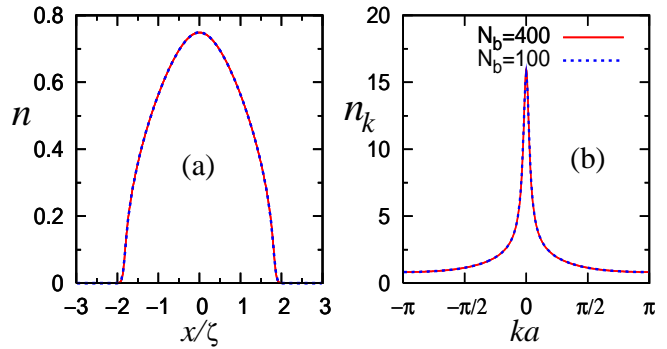}
\end{center} \vspace{-0.6cm}
\caption{(Color online) Density and momentum profiles
in two traps with different filling, 
but the same $\tilde{\rho}=2$ and $k_B T=0.1 t$.}
\label{PerfilScaling}
\end{figure}

To conclude this section we show in Fig.\ \ref{PerfilScaling} 
a comparison between density and momentum profiles for 
100 and 400 HCB's at $k_B T=0.1t$ and $\tilde{\rho}=2$. 
While even at zero temperature the density profiles do not differ 
\cite{rigol04_1,rigol03_3}, the ground-state peak $n_{k=0}$ 
would have been 2 times larger for 400 HCB's than for 100 HCB's. 
At $k_B T=0.1 t$ both momentum distribution functions are 
almost indistinguishable.

\section{Grand-canonical vs canonical ensemble}

In the previous sections we have discussed effects of the 
temperature on trapped HCB's in 1D. The starting point for our 
calculations was the grand-canonical ensemble, in which 
the system is assumed to be in thermal equilibrium with 
a large reservoir with temperature $T$ and chemical 
potential $\mu$. The chemical potential was then chosen 
to obtain the desired average number of particles in the 
trap. In this section we analyze the changes introduced 
by the grand-canonical fluctuations of the particle number 
with respect to a fixed-$N_b$ canonical description, which 
may be more relevant to describe trapped ultracold quantum gases 
where no particle reservoir is available.

In the thermodynamic limit both descriptions are known to 
provide the same predictions \cite{huang87}. On the other 
hand, for noninteracting bosonic systems with a {\em mesoscopic} 
number of particles, it has been shown that the differences
between the grand-canonical and canonical condensate fractions
can be as large as $10\%$ (for $N_b=100$) close to the
BEC transition point and decreasing logarithmically with 
increasing number of particles. In the present work we 
have been dealing with the opposite case---i.e., infinite repulsion. 
Interactions are known to suppress fluctuations of the number 
of particles in the grand-canonical ensemble \cite{huang87}, 
but since recent experiments with HCB's on optical lattices 
achieved only up to 20 HCB's in around 50 lattice sites, it 
is useful to present an estimate of the difference between 
both ensembles for such small systems.  
 
In order to obtain the canonical one-particle density matrix  
we use the ground-state approach of Ref.\ \cite{rigol04_1}. 
We calculate the Green's function of all states 
$|\Psi^{n}_{HCB}\rangle$ with $N_b$ bosons in $N$ lattice 
sites---i.e., of $N_s=N!/(N-N_b)!N_b!$ states. The canonical Green's 
function at temperature $T$ is obtained as the sum
\begin{equation}
\label{green3} G^C_{ij}=\dfrac{1}{Z^C}\sum_{n=1}^{N_s} e^{-E_n/k_B T}
\langle \Psi^{n}_{HCB}|b^{}_{i}b^\dagger_{j}|\Psi^{n}_{HCB}\rangle,
\end{equation}
where $e^{-E_n/k_B T}$ ($E_n$ is the energy of state 
$|\Psi^{n}_{HCB}\rangle$) is the Boltzmann factor and 
$Z^C$ the canonical partition function:
\begin{equation}
Z^C=\sum_{n=1}^{N_s} e^{-E_n/k_B T}.
\end{equation}
The canonical one-particle density matrix is then
\begin{eqnarray}
\rho^C_{ij}&=& \dfrac{1}{Z^C}\sum_{n=1}^{N_s} e^{-E_n/k_B T}
\langle \Psi^{n}_{HCB}|b^\dagger_{i}b^{}_{j}|\Psi^{n}_{HCB}\rangle 
\nonumber \\ 
&=&G^C_{ij}+\delta_{ij}\left(1-2 G^C_{ii} \right).
\end{eqnarray}

\begin{figure}[b]
\begin{center}
\includegraphics[width=0.48\textwidth,height=0.28\textwidth]
{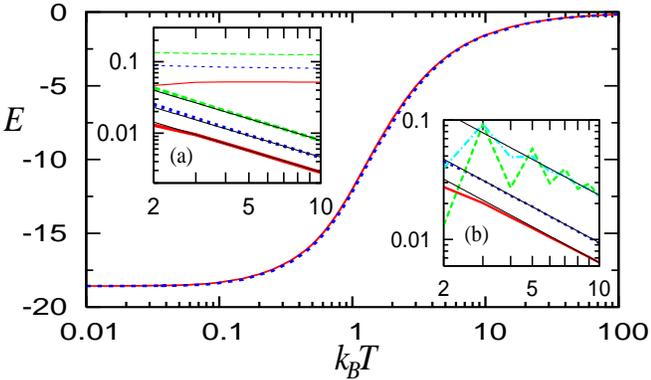}
\end{center} \vspace{-0.6cm}
\caption{(Color online) Energy (in units of $t$) vs temperature 
(also in units of $t$) for a box with 
50 lattice sites and 10 particles. The grand-canonical ($E$) and 
canonical ($E^C$) results are plotted as solid and dotted lines, 
respectively. Inset (a) shows $E-E^C$ (thin lines) and 
$\delta E=(E-E^C)/|E^C|$
(thick lines) vs $N_b$ for $k_B T=0.1t$ (solid line), 
$0.2t$ (dotted line), $0.5t$ (dashed line). 
Thin solid lines following $\delta E$ 
exhibit $1/N_b$ behavior. Inset (b) depicts $\delta n_k$ for HCB's
and $\delta n^f_k$ for fermions vs $N_b$ (see text). 
The temperatures are in this case $k_B T=0.1t$ (solid line
for $\delta n_k$, dashed line for $\delta n^f_k$) and 
$k_B T=0.2t$ (dotted line for $\delta n_k$, dash-dotted 
line for $\delta n^f_k$). Thin solid lines following 
these results exhibit $1/N_b$ behavior. In both insets the 
density $\rho=N_b/N=0.2$ was kept constant when changing
the number of particles.}
\label{EvsT_CvsGCBOX}
\end{figure}
In Fig.\ \ref{EvsT_CvsGCBOX} we show results obtained for the 
grand-canonical ($E$) and canonical ($E^C$) energies of 10 HCB's 
in a box with 50 lattice sites as a function of the temperature. 
At the scale of the figure they are indistinguishable.
More information can be obtained in the inset (a) where we plot as 
thin lines the energy difference between both ensembles as a 
function of the number of particles in boxes with densities 
$N_b/N=0.2$. As seen in this inset even for such small systems 
the difference almost does not change with $N_b$, and it is 
always smaller than the energy unit $t$. Considering that the 
modulus of the energy increases linearly with the system size, 
the relative difference between both ensembles decreases 
$\delta E\equiv (E-E^C)/|E^C|\sim 1/N_b$ [thick lines in inset (a)]. 
For $N_b=10$ and $N=50$ one can see that $\delta E$ is below $1\%$ 
for temperatures up to $k_B T=0.5t$.   
As the temperature increases beyond $k_B T=t$ the differences 
between $E$ and $E^C$ start to decrease, which together with
the decrease of the modulus of $E^C$ shown in 
Fig.\ \ref{EvsT_CvsGCBOX} produces a saturation of $\delta E$ 
at around $2\%$ for $k_B T>10t$. Then for $N_b=10$ and $N=50$ 
the maximum $\delta E$ is just a $2\%$ of the energy. We have 
also studied other densities keeping $N=50$, and the results 
obtained for the maximum $\delta E$ were exactly the same $2\%$.

While Kinoshita {\it et~al.} \cite{kinoshita04} used the energy 
of the system to confirm the achievement of the hard-core limit, 
Paredes {\it et~al.} \cite{paredes04} considered the momentum 
distribution function. In the inset (b) of Fig.\ \ref{EvsT_CvsGCBOX} 
we show the relative difference 
[$\delta n_k\equiv(\sum_k |n_k-n^C_k|)/(\sum_k n^C_k)$] between 
the grand-canonical $n_k$ and canonical $n^C_k$ calculation 
of the momentum distribution function.
The relative differences for $n_k$ although larger than the 
corresponding ones for the energy are still small and 
also reduce $\sim 1/N_b$ with increasing the system size. 
For $N_b=10$ and $N=50$ they are smaller than 
$1\%$ up to $k_B T=0.2t$.

It is also useful to calculate the differences between
the grand-canonical and canonical ensemble for the equivalent 
noninteracting fermions. This may be relevant for systems 
like the ones recently achieved experimentally by 
K\"ohl {\it et~al.}\ \cite{kohl05}. For noninteracting fermions, 
the energy differences between both ensembles are identical 
to the ones of the HCB's due to the mapping,  
Eqs.\ (\ref{HamHCB})--(\ref{HamFerm}), 
so that as shown in Fig.\ \ref{EvsT_CvsGCBOX} and its inset (a) 
they are small. For the fermionic momentum distribution 
function the HCB results do not apply. We have also calculated 
the fermionic relative difference 
[$\delta n^f_k\equiv(\sum_k |n^f_k-n^{f,C}_k|)/(\sum_k n^{f,C}_k)$] 
between the grand-canonical $n_k^f$ and canonical $n^{f,C}_k$ 
calculation of $n_k$. They are larger than for the ones 
of the HCB's as shown in the inset (b) 
of Fig.\ \ref{EvsT_CvsGCBOX}. However, they are still small
for the experimentally accessible system sizes. For $N_f=10$ 
and $N=50$ they are smaller than $3\%$ for $k_B T=0.2t$.  
Apart from an even-odd effect that decreases with increasing
the temperature, $\delta n^f_k$ also decreases $\sim 1/N_b$ 
with increasing the system size.

\begin{figure}[h]
\begin{center}
\includegraphics[width=0.48\textwidth,height=0.28\textwidth]
{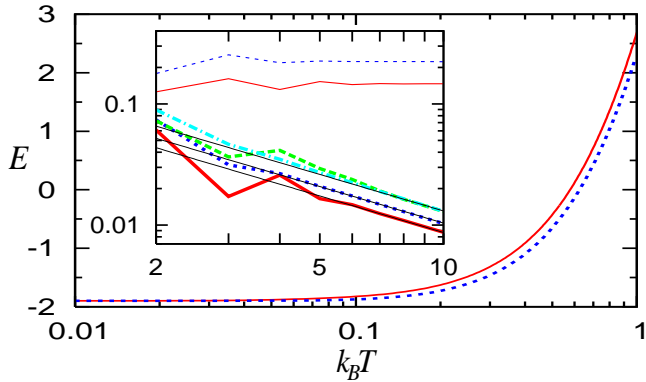}
\end{center} \vspace{-0.6cm}
\caption{(Color online) Energy vs temperature 
for a harmonic trap with 10 particles and $\tilde{\rho}=2$. 
The grand-canonical ($E$) and canonical ($E^C$) results 
are plotted as solid and dotted lines, respectively.
The inset shows $E-E^C$ (thin lines) and $\delta n_k$ 
(thick lines) vs $N_b$ for $k_B T=0.3t$ (solid line) 
and $0.5t$ (dotted line). 
The $\delta n^f_k$ of the fermions is also shown as a 
function of $N_f$ (thick lines) for $k_B T=0.3t$ 
(dashed line) and $0.5t$ (dash-dotted line).
Thin solid lines following the $\delta n_k$ 
and $\delta n^f_k$ results exhibit $1/N_b$ 
behavior. In the inset the characteristic density 
$\tilde{\rho}=2$ was kept constant when changing 
the number of particles.}
\label{EvsT_CvsGCTRAP}
\end{figure}

The introduction of a harmonic trap does not (qualitatively) 
change the results obtained in a box. 
In Fig.\ \ref{EvsT_CvsGCTRAP} we show the 
grand-canonical and canonical results of the energy in a 
harmonic trap with 10 particles and $\tilde{\rho}=2$ 
as a function of the temperature. Contrary to the box, 
in a harmonic trap the energy is not bounded from above 
for very large temperatures. This is because the HCB cloud can 
increase its size and consequently its potential energy. The 
energy differences between both ensembles, when changing the 
number of particles keeping $\tilde{\rho}=2$ constant, 
are shown in the inset of Fig.\ \ref{EvsT_CvsGCTRAP}. As for 
the box they are almost independent of $N_b$, 
and smaller than $t$. The results for the relative differences 
between the momentum distribution functions for HCB's and 
noninteracting fermions in both ensembles are also shown 
in the inset. Their behavior is very similar to the one of 
the box in inset (b) of Fig.\ \ref{EvsT_CvsGCBOX}.

As mentioned before, Herzog and Olshanii \cite{herzog97} 
discussed the grand canonical and canonical differences 
between the condensate fraction for noninteracting bosons 
in harmonic traps. At finite repulsive interactions, in 1D, 
there is no BEC even at zero temperature. Still, for the 
HCB's we have calculated the differences between the 
largest eigenvalue of the one-particle density matrix 
(equivalent to the condensate 
occupation for BEC \cite{leggett01}) in the grand-canonical 
and canonical ensembles. As for $\delta n_k$ in the inset of 
Fig.\ \ref{EvsT_CvsGCTRAP}, we find that the difference 
between them decreases $\sim 1/N_b$ with increasing 
number of particles in the system. This clearly contrasts 
with the $\sim 1/\ln(N_b)$ obtained for the noninteracting 
case \cite{herzog97}.

\begin{figure}[h]
\begin{center}
\includegraphics[width=0.45\textwidth]
{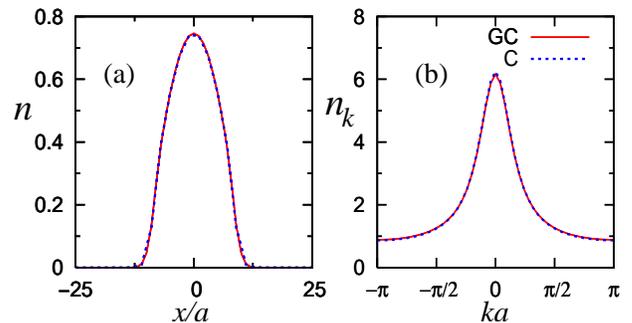}
\end{center} \vspace{-0.7cm}
\caption{(Color online) Density and momentum profiles of
10 HCB's in a harmonic trap with $\tilde{\rho}=2$ at $k_B T=0.5 t$.
The grand-canonical and canonical results are plotted as solid 
and dashed lines respectively.}
\label{PerfilGCvsC}
\end{figure}

We conclude by explicitly showing in 
Fig.\ \ref{PerfilGCvsC} the density profiles and 
momentum distribution functions of 10 HCB's 
in a harmonic trap with $\tilde{\rho}=2$ at $k_B T=0.5t$ 
as obtained from the grand-canonical and canonical descriptions.
They are basically indistinguishable. Then, even for the small 
system sizes achieved experimentally, one can rely on the 
grand-canonical description for strongly correlated HCB's 
for the physical quantities described here. 
The same conclusion applies to noninteracting fermions.

\section{Conclusions}

We have presented an exact study of finite-temperature properties
of HCB's confined on 1D lattices. In order to solve this problem we 
have used the Jordan-Wigner transformation to map HCB's into 
noninteracting fermions. After the mapping, properties of Slater 
determinants allowed us to obtain an exact expression for the HCB 
one-particle density matrix in terms of determinants of 
$N\times N$ matrices, which are evaluated numerically. Our approach 
represents an alternative for finite systems to previous works 
that considered the thermodynamic limit
\cite{lieb61,mccoy68,vaidya78,mccoy83,tonegawa81} for periodic
and open chains and in which Toeplitz determinants were involved. 

We have shown that the effects of small finite temperatures are 
very important when dealing with quantities related to off-diagonal
one-particle correlations like the momentum distribution function
and the natural orbitals. These finite-temperature effects depend 
strongly on the system size. On the other hand, observables related to 
diagonal one-particle correlations (identical for fermions), 
like density profiles, are much less affected at low temperatures.  
Explicit results for the behavior of 
all these quantities versus temperature were given for system sizes 
that range from the ones recently achieved experimentally up to 
20 times larger.

Finally, we have compared grand-canonical and canonical results
for energies and momentum distribution functions of HCB's and 
noninteracting fermions for small systems, like the ones achieved 
experimentally. In spite of the mesoscopic number of particles
we have shown that for these system sizes the effects of the 
grand-canonical fluctuations of the particle number are 
very small and one can rely on a grand-canonical approach.

Although all our calculations are exact for infinite on-site 
$U$ repulsion, for very strong but finite $U$ our conclusions 
are still valid. In this case $1/U$ acts like a perturbation 
to the noninteracting spinless fermion Hamiltonian 
\cite{cazalilla03}. Rey {\it et~al.} \cite{rey05} have 
recently presented results obtained by exact diagonalization 
that support the above conclusion \cite{rey05}. A connection 
to experimentally relevant parameters can be also found in 
Ref.\ \cite{rey05}.

\begin{acknowledgments}

We are grateful to G. G. Batrouni, A. Muramatsu, M. Olshanii, 
R. T. Scalettar, and R. R. P. Singh for stimulating discussions and
comments on the manuscript and to M. Arikawa for 
pointing out several references. This work was supported 
by Grant Nos.\ NSF-DMR-0312261, NSF-DMR-0240918, 
and NSF-ITR-0313390.

\end{acknowledgments}


\begin{thebibliography}{99}
\bibitem{BW} Or what is the same, for energies exceeding in $4t$ 
(which is the bandwidth in 1D) the energy of the lowest eigenstate 
of the Hamiltonian \cite{rigol03_3}.
\end{thebibliography}
\end{document}